\documentclass[12pt]{article}
\usepackage{amssymb}
\usepackage{amsmath}
\usepackage{color}
\usepackage{euscript}
\def\d{\delta}

\def\be{\begin{equation}}
\def\ee{\end{equation}}
\def\arr{\begin{array}{rll}}
\def\ea{\end{array}}
\def\bea{\begin{eqnarray}}
\def\eea{\end{eqnarray}}

\def\N2{$N{=}2$}

\def\>{\rangle}
\def\<{\langle}
\def\+{\dagger}
\def\={\ =\ }

\begin{document}
\renewcommand{\thefootnote}{\arabic{footnote}}
\noindent
\begin{titlepage}
\setcounter{page}{0}
\begin{flushright}
$\qquad$
\end{flushright}
\vskip 3cm
\begin{center}
{\LARGE\bf{Remark on higher-derivative mechanics}
\vskip 0.5cm
{\LARGE\bf with $l$-conformal Galilei symmetry}}
\vskip 1cm
$
\textrm{\Large Ivan Masterov\ }
$
\vskip 0.7cm
{\it
Laboratory of Mathematical Physics, Tomsk Polytechnic University, \\
634050 Tomsk, Lenin Ave. 30, Russian Federation}
\vskip 0.7cm
{E-mail: masterov@tpu.ru}

\end{center}
\vskip 1cm
\begin{abstract}
\noindent
Based on the results in [Nucl. Phys. B 866 (2013) 212], we consider a way to construct a higher-derivative mechanical model which possesses the $l$-conformal Galilei symmetry. The dynamical system describes generalized Pais-Uhlenbeck oscillator in the presence of an effective external field.
\end{abstract}


\vskip 0.5cm


\end{titlepage}

\noindent
{\bf{\large 1. Introduction}}
\vskip 0.5cm
\noindent
The concept of a symmetry plays an important role in modern theoretical physics. In general, there are two independent ways of dealing with symmetries in mechanics. On the one hand, we may find an invariance group for a given model. On the other hand, one constructs new systems as dynamical realizations of a concrete Lie algebra. In this context, the so-called $l$-conformal Galilei algebra \cite{Henkel}-\cite{Negro_1}, where $l$ being positive integer or half-integer, has recently attracted considerable attention (see, e.g., \cite{Henkel2002}-\cite{C} and references therein).

For $l>1/2$, the $l$-conformal Galilei algebra includes generators of accelerations. Because of this, dynamical realizations of this algebra in general involve higher-derivative terms \cite{Gomis}-\cite{AG}, \cite{AGKM} (see also \cite{DH}). For $l=1$, the way to realize the algebra in a system without higher derivatives has been introduced in Ref. \cite{FIL}. It has been shown that the system of equations
\bea\label{1}
\mbox{(a)}\;\,\ddot{\rho}=\frac{\gamma^2}{\rho^3},\qquad\mbox{(b)}\;\, \mathbb{D}^2\chi_i+4\gamma^2\chi_i=0,\quad\mbox{where}\quad\mathbb{D}=\rho^2\frac{d}{dt},
\eea
enjoys such a symmetry. Moreover, as was shown in Ref. \cite{GM}, this system may accommodate the $l$-conformal Galilei symmetry for any positive integer $l$. A specific feature of the realization (\ref{1}) is that the generators of accelerations are functionally dependent on those related with other symmetries \cite{GM}.

According to the analysis in Ref. \cite{GM}, the equation (\ref{1}a) can be interpreted as describing an effective external field, while the oscillator-like equation (\ref{1}b) describes a particle moving in this field. As was shown in \cite{GM} (see also the related discussion in Ref. \cite{Galajinsky_1}), a motion of this particle is bounded and therefore is stable.

The classical stability is fundamentally important issue for higher-derivative mechanical systems which in general reveal instability in classical dynamics \cite{Ostrogradski}. At the same time, there exist higher-derivative models which are stable. An example of such systems is the celebrated Pais-Uhlenbeck (PU) oscillator \cite{Pais}. For the case of distinct frequencies, the motion of this model is bounded and small perturbations of initial conditions do not affect this feature. A trouble with runaway solutions may appear in the case of interacting PU oscillator \cite{Nesterenko} (for a detailed discussion see, e.g., Ref. \cite{Woodard}). This motivates a development of methods to construct stable deformations and generalizations of the original PU oscillator \cite{Smilga_1}-\cite{Alt_1}. The purpose of this work is to construct a stable generalization of the PU oscillator, which accommodates the $l$-conformal Galilei symmetry, by using the results in \cite{GM}. For $l=2$, this will be done in the next section while the case of general $l$ is treated in Sect. 3. We summarize our results and discuss further possible developments in concluding Sect. 4. Throughout the work summation over repeated spatial indices is understood. A number of dots over the fields designate the number of derivatives with respect to time.

\vskip 0.5cm
\noindent
{\bf{\large 2. The $l=2$-conformal Galilei algebra}}
\vskip 0.5cm

The $l$-conformal Galilei algebra involves the generator of time translations $H$, the generators of space rotations $M_{ij}$, a chain of vector generators $C_i^{(n)}$ with $n=0,1,..,2l$ as well as those of dilatations $D$ and special conformal transformations $K$. The structure relations in the algebra read
\bea\label{algebra}
\begin{aligned}
&
[H,D]=H,&&[H,C_i^{(n)}]=nC_i^{(n-1)},
\\[2pt]
&
[H,K]=2D,&&[D,K]=K,
\\[2pt]
&
[D,C_i^{(n)}]=(n-l)C_i^{(n)},&&[K,C_i^{(n)}]=(n-2l)C_i^{(n+1)},
\\[2pt]
&
[M_{ij},C_k^{(n)}]=-\d_{ik}C_j^{(n)}+\d_{jk}C_i^{(n)},&&[M_{ij},M_{kl}]=-\d_{ik}M_{jl}-\d_{jl}M_{ik}+\d_{il}M_{jk}+\d_{jk}M_{il}.
\end{aligned}
\eea
The generators $C_i^{(0)}$ and $C_i^{(1)}$ produce space translations and Galilei boosts, respectively, while other vector generators correspond to accelerations.

For the purpose of the present work, let us consider the $l=2$-conformal Galilei algebra in more details. As was shown in Ref. \cite{GM}, the following equations
\bea\label{equ}
\mathbb{D}\xi_{0,i}=0,\quad \mathbb{D}^2\xi_{1,i}+4\gamma^2\xi_{1,i}=0,\quad \mathbb{D}^2\xi_{2,i}+16\gamma^2\xi_{2,i}=0,
\eea
together with (\ref{1}a), may accommodate such a symmetry. On the space of fields $\rho$, $\xi_{k,i}$ the $l=2$-conformal Galilei group acts as follows:
\bea\label{transform}
&&
\qquad\qquad\qquad{\rho}'(t)=\rho(t)+\frac{1}{2}(c+2bt)\rho(t)-(a+bt^2+ct)\dot{\rho}(t),
\\[2pt]
&&
{\xi'}_{k,i}(t)=\xi_{k,i}(t)+\sum_{n=0}^{4}\sum_{s=0}^{n}\frac{n!}{s!(n-s)!}t^{n-s}u_{k,s}\lambda_{n,i}-(a+bt^2+ct)\dot{\xi}_{k,i}(t)-\omega_{ij}\xi_{k,j}(t),
\eea
where $a$, $b$, $c$, $\lambda_{n,i}$, and $\omega_{ij}$ are infinitesimal parameters corresponding to time translations, special conformal transformations, dilatations, vector generators, and rotations, respectively. The functions $u_{k,s}$ are defined by
\bea\label{Table}
\begin{array}{|c|c|c|c|c|c|}
\hline
s&\quad\qquad0\qquad\quad&\quad\qquad1\qquad\quad&\quad\qquad2\qquad\quad&\quad\qquad3\qquad\quad&\quad\qquad4\qquad\quad
\\[2pt]
\hline
& & & & &\\
u_{0,s}&{\displaystyle\frac{3\gamma^2}{\rho^4}+\frac{6\dot{\rho}^2}{\rho^2}+\frac{3\dot{\rho}^4}{\gamma^2}}& {\displaystyle-\frac{3\dot{\rho}}{\rho}-\frac{3\rho\dot{\rho}^3}{\gamma^2}}&
{\displaystyle\frac{3\rho^2\dot{\rho}^2}{\gamma^2}+1}&{\displaystyle-\frac{3\rho^3\dot{\rho}}{\gamma^2}}&
{\displaystyle\frac{3\rho^4}{\gamma^2}}
\\[2pt]
& & & & &\\
\hline
&&&&&\\
u_{1,s}&{\displaystyle\frac{\gamma^2}{\rho^4}-\frac{\dot{\rho}^4}{\gamma^2}}&{\displaystyle\frac{\rho\dot{\rho}^3}{\gamma^2}}&
{\displaystyle-\frac{\rho^2\dot{\rho}^2}{\gamma^2}}&{\displaystyle\frac{\rho^3\dot{\rho}}{\gamma^2}}&{\displaystyle-\frac{\rho^4}{\gamma^2}}
\\[2pt]
&&&&&\\
\hline
&&&&&\\
u_{2,s}&{\displaystyle\frac{\gamma^2}{\rho^4}+\frac{\dot{\rho}^4}{\gamma^2}-\frac{6\dot{\rho}^2}{\rho^2}}&
{\displaystyle-\frac{\rho\dot{\rho}^3}{\gamma^2}+\frac{3\dot{\rho}}{\rho}}&{\displaystyle\frac{\rho^2\dot{\rho}^2}{\gamma^2}-1}&
{\displaystyle-\frac{\rho^3\dot{\rho}}{\gamma^2}}&{\displaystyle\frac{\rho^4}{\gamma^2}}
\\[2pt]
&&&&&\\
\hline
\end{array}
\nonumber
\eea
It is straightforward to verify that commutators $[\delta_1,\delta_2]$ of variations $\d\rho(t)=\rho'(t)-\rho(t)$, $\d\xi_{k,i}(t)={\xi'}_{k,i}(t)-\xi_{k,i}(t)$ reproduce the algebra (\ref{algebra}) for $l=2$ \cite{GM}. The form of the transformations (\ref{transform}) shows that the $l=2$-conformal Galilei algebra can be realized in a system which consists of two equations. The first one is (\ref{1}a), while any equation in (\ref{equ}) can play a role of the second equation of the system.

Let us consider the variable $\chi_i=\xi_{1,i}+\xi_{2,i}$ whose dynamics is governed by the following equation of motion:
\bea\label{PU4}
\mathbb{D}^4\chi_i+20\gamma^2\mathbb{D}^2\chi_i+64\gamma^4\chi_i=0.
\eea
It is evident that the system of equations (\ref{1}a), (\ref{PU4}) also accommodates the $l=2$-conformal Galilei symmetry. The corresponding symmetry transformations involve (\ref{transform}) and
\bea\label{transform1}
{\chi'}_{i}(t)=\chi_{i}(t)+\sum_{n=0}^{4}\sum_{s=0}^{n}\frac{n!}{s!(n-s)!}t^{n-s}(u_{1,s}+u_{2,s})\lambda_{n,i}-(a+bt^2+ct)\dot{\chi}_{i}(t)-
\omega_{ij}\chi_{j}(t).
\eea

The system of equations (\ref{1}a), (\ref{PU4}) is not Lagrangian. But according to the analysis in Ref. \cite{GM}, one may attempt to obtain conserved charges associated with the transformations (\ref{transform}), (\ref{transform1}) by applying the Noether theorem to some effective actions. In particular, one may use the action functional
\bea
S=\frac{1}{2}\int dt\left(\dot{\rho}^2-\frac{\gamma^2}{\rho^2}\right)
\nonumber
\eea
to derive the following conserved quantities for the system (\ref{1}a), (\ref{PU4})
\bea
\mathcal{H}=\dot{\rho}^2+\frac{\gamma^2}{\rho^2},\qquad\mathcal{D}=\rho\dot{\rho}-t\mathcal{H},\qquad \mathcal{K}=t^2\mathcal{H}-2t\rho\dot{\rho}+\rho^2,
\nonumber
\eea
which correspond to invariance of this action under the time translations, dilatations, and special conformal transformations, respectively \cite{Alfaro}.
At the same time, the invariance of the action functional
\bea
S=\frac{1}{2}\int \frac{dt}{\rho^2}\chi_i\left(\mathbb{D}^4\chi_i+20\gamma^2\mathbb{D}^2\chi_i+64\gamma^4\chi_i\right)
\nonumber
\eea
under the transformations
\bea\label{tr123}
{\chi'}_{i}(t)=\chi_{i}(t)+\sum_{n=0}^{4}\sum_{s=0}^{n}\frac{n!}{s!(n-s)!}t^{n-s}(u_{1,s}+u_{2,s})\lambda_{n,i}
\eea
leads to the conservation of the following quantities
\bea
\begin{array}{c}
\mathcal{C}_i^{(0)}=2(3\rho^4\dot{\rho}^2-\gamma^2\rho^2)\dddot{\chi}_i+16(3\rho^3\dot{\rho}^3-2\gamma^2\rho\dot{\rho})\ddot{\chi}_i
+8\left(9\rho^2\dot{\rho}^4-2\gamma^2\dot{\rho}^2-{\displaystyle \frac{3\gamma^4}{\rho^{2}}}\right)\dot{\chi}_i-{\displaystyle \frac{128\gamma^4\dot{\rho}}{\rho^{3}}}\chi_i,
\\[2pt]
\mathcal{C}_i^{(1)}=t \mathcal{C}_i^{(0)}-3\rho^5\dot{\rho}\dddot{\chi}_i-(27\rho^4\dot{\rho}^2-
5\gamma^2\rho^2)\ddot{\chi}_i-8(6\rho^3\dot{\rho}^3+\gamma^2\rho\dot{\rho})\dot{\chi}_i+{\displaystyle\frac{32\gamma^4}{\rho^{2}}}\chi_i,
\\[6pt]
\mathcal{C}_i^{(2)}=-t^2 \mathcal{C}_i^{(0)}+2t \mathcal{C}_i^{(1)}+\rho^6\dddot{\chi}_i+12\rho^5\dot{\rho}\ddot{\chi}_i+
6(5\rho^4\dot{\rho}^2+\gamma^2\rho^2)\dot{\chi}_i,
\end{array}
\nonumber
\eea
\bea
\begin{array}{c}
\mathcal{C}_i^{(3)}=t^3 \mathcal{C}_i^{(0)}-3t^2 \mathcal{C}_i^{(1)}+3t \mathcal{C}_i^{(2)}-3\rho^6\ddot{\chi}_i-18\rho^5\dot{\rho}\dot{\chi}_i,
\\[6pt]
\mathcal{C}_i^{(4)}=-t^4 \mathcal{C}_i^{(0)}+4t^3 \mathcal{C}_i^{(1)}-6t^2 \mathcal{C}_i^{(2)}+4t \mathcal{C}_i^{(3)}+12\rho^6\dot{\chi}_i.
\end{array}
\nonumber
\eea
When these constants of motion are derived, $\rho$ is to be treated as external field obeying the equation (\ref{1}a). These integrals of motion prove to be functionally dependent
\bea
&&
C_i^{(4)}=-\frac{(\mathcal{D}^2+\gamma^2)^2}{\mathcal{H}^4}C_i^{(0)}-\frac{4\mathcal{D}(\mathcal{D}^2+\gamma^2)}{\mathcal{H}^3}C_i^{(1)}-
\frac{2(3\mathcal{D}^2+\gamma^2)}{\mathcal{H}^2}C_i^{(2)}-\frac{4\mathcal{D}}{\mathcal{H}}C_i^{(3)}.
\nonumber
\eea
This correlates with the fact that four functionally independent integrals of motion are enough to integrate the fourth-order differential equation (see also the discussion in Ref. \cite{GM}).

The equation (\ref{PU4}) can be presented in the following form:
\bea\label{schi}
\frac{d^4\chi_i}{ds^4}+20\gamma^2\frac{d^2\chi_i}{ds^2}+64\gamma^4\chi_i=0
\eea
with the aid of the function
\bea
s(t)=\frac{1}{\gamma}\arctan{\left(\frac{\mathcal{D}+t\mathcal{H}}{\gamma}\right)},\quad \mathbb{D}=\frac{d}{ds}.
\nonumber
\eea
At first sight it may appear that the change of time $t\rightarrow s(t)$ allows one to discard the field $\rho$ from our consideration, i.e. that the equation (\ref{schi}) itself accommodates the $l$-conformal Galilei symmetry. This is not true because the field $\rho$ is a principal ingredient of the symmetry transformations (\ref{tr123}). In the framework of the present study we cannot produce another transformations which do not involve $\rho$.

The general solution of the equation (\ref{PU4}) can be presented as follows
\bea\label{sol1}
\chi_i(t)=A_i\cos{(2\gamma s(t))}+B_i\sin{(2\gamma s(t))}+C_i\cos{(4\gamma s(t))}+D_i\sin{(4\gamma s(t))},
\eea
where $A_i$, $B_i$, $C_i$, and $D_i$ are constants of integration. The function $s(t)$ is defined by \cite{GM}
\bea
s(t)=\frac{1}{\gamma}\arctan{\left(\frac{\mathcal{D}+t\mathcal{H}}{\gamma}\right)},\quad \mathbb{D}=\frac{d}{ds}.
\nonumber
\eea
So, the motion of the particle (\ref{PU4}) is bounded and is stable.

The equation (\ref{PU4}) is similar to the equation of motion of the ordinary fourth-order PU oscillator when frequencies of oscillation form ratio $\frac{\omega_1}{\omega_2}=\frac{1}{2}$
\bea\label{PUO}
x_i^{(4)}+20\gamma^2\ddot{x}_i+64\gamma^4 x_i=0.
\eea
In this sense the model (\ref{PU4}) can be viewed as a generalization of the PU oscillator (\ref{PUO}) (about conformal invariance of the ordinary PU oscillator see \cite{Galajinsky_1}, \cite{PU}-\cite{Andr}). Let us confront qualitative behavior of the particle described by (\ref{PU4}) with that of the system (\ref{PUO}). To this end, we write a general solution of (\ref{PUO}) with the same constants of integration as in (\ref{sol1})
\bea\label{sol2}
x_i(t)=A_i\cos{(2\gamma t)}+B_i\sin{(2\gamma t)}+C_i\cos{(4\gamma t)}+D_i\sin{(4\gamma t)}.
\eea
For definiteness, we assume $\gamma$ to be positive. According to the solution (\ref{sol2}), the PU oscillator makes a cyclic motion with the period of oscillations $T=\frac{\pi}{\gamma}$ along a curve whose form is defined by initial conditions. A trajectory of the particle (\ref{PU4}) is the same curve with one point corresponding to the polar angle $\phi=2\gamma s=\pi$ removed. Being initially at rest close to $\chi_i=C_i-A_i$ (as $t\rightarrow -\infty$), the particle (\ref{PU4}) starts moving along a curve. When one cycle is finished, the particle freezes up at the initial point as $t\rightarrow\infty$.

The first-order equation in (\ref{equ}) can be used to construct the third-order dynamical system. Indeed, the dynamics of the variables $\psi_{i}^k=\xi_{0,i}+\xi_{k,i}$ with $k=1,2$ is governed by
\bea\label{PU3}
\mathbb{D}^3\psi_{i}^{k}+(2k\gamma)^2\mathbb{D}\psi_{i}^{k}=0.
\eea
These equations can be viewed as describing a generalization of the third-order PU oscillator (see, e.g., \cite{Alt_1} and references therein). The system of equations (\ref{1}a) and (\ref{PU3}) also possesses the $l=2$-conformal Galilei symmetry. When $i=1,2$, effective actions for these equations can be defined as follows:
\bea
S=\frac{1}{2}\int\frac{dt}{\rho^2}\epsilon_{ij}\psi_{i}^k(\mathbb{D}^3\psi_{j}^{k}+(2k\gamma)^2\mathbb{D}\psi_{j}^k),
\nonumber
\eea
where $\epsilon_{ij}$ is the Levi-Civit\'{a} symbol.
Then, the integrals of motion which correspond to vector generators in the algebra (\ref{algebra}) can be derived via the Noether theorem. For example, for $k=1$ these have the form
\bea
\begin{array}{c}
\mathcal{C}_i^{(0)}=-2\epsilon_{ij}\left[\left(\frac{\rho^4\dot{\rho}^4}{\gamma^2}+3\rho^2\dot{\rho}^2+
2\gamma^2\right)\ddot{\psi}_{j}^1+2\left(\frac{\rho^3\dot{\rho}^5}{\gamma^2}+4\rho\dot{\rho}^3+
\frac{3\gamma^2\dot{\rho}}{\rho}\right)\dot{\psi}_{j}^{1}+6\left(\dot{\rho}^4+\frac{2\gamma^2\dot{\rho}^2}{\rho^2}+
\frac{\gamma^4}{\rho^4}\right)\psi_{j}^1\right],
\\[8pt]
\mathcal{C}_i^{(1)}=t\mathcal{C}_i^{(0)}+\epsilon_{ij}\left[\left(3\rho^3\dot{\rho}+\frac{2\rho^5\dot{\rho}^3}{\gamma^2}\right)\ddot{\psi}^{1}_{j}+
\left(\gamma^2+9\rho^2\dot{\rho}^2+\frac{4\rho^4\dot{\rho}^4}{\gamma^2}\right)\dot{\psi}^{1}_{j}+\left(12\rho\dot{\rho}^3+
\frac{12\gamma^2\dot{\rho}}{\rho}\right)\psi^{1}_{i}\right],
\\[8pt]
\mathcal{C}_i^{(2)}=-t^2\mathcal{C}_i^{(0)}+2t\mathcal{C}_i^{(1)}-\epsilon_{ij}\left[\left(\rho^4+\frac{2\rho^6\dot{\rho}^2}{\gamma^2}\right)\ddot{\psi}^1_j+
4\left(\rho^3\dot{\rho}+\frac{\rho^5\dot{\rho}^3}{\gamma^2}\right)\dot{\psi}_j^1+4\left(\gamma^2+3\rho^2\dot{\rho}^2\right)\psi_{j}^1\right],
\\[8pt]
\mathcal{C}_i^{(3)}={\displaystyle-\frac{\mathcal{D}(\mathcal{D}^2-\gamma^2)}{\mathcal{H}^3}\mathcal{C}_i^{(0)}+
\frac{\gamma^2-3\mathcal{D}^2}{\mathcal{H}^2}\mathcal{C}_i^{(1)}-
\frac{3\mathcal{D}}{\mathcal{H}}\mathcal{C}_i^{(2)}},
\nonumber
\\[8pt]
\mathcal{C}_i^{(4)}={\displaystyle\frac{3\mathcal{D}^4+2\mathcal{D}^2\gamma^2-\gamma^4}{\mathcal{H}^4}\mathcal{C}_i^{(0)}+
\frac{8\mathcal{D}(\mathcal{D}^2+\gamma^2)}{\mathcal{H}^3}\mathcal{C}_i^{(1)}+
\frac{6(\mathcal{D}^2+\gamma^2)}{\mathcal{H}^2}\mathcal{C}_i^{(2)}}.
\end{array}
\nonumber
\eea
So, for generalized third-order PU oscillator, only three vector integrals of motion are functionally independent as it should be.

\vskip 0.5cm
\noindent
{\bf{\large 3. The case of arbitrary $l$}}
\vskip 0.5cm

Let us turn to the case of arbitrary $l$. As was shown in Ref. \cite{GM}, the system of equations, which involves (\ref{1}a) and
\bea\label{equ1}
\mathbb{D}x_{n,i}=\sum_{m=0}^{2l}x_{m,i}A^{mn},\qquad n=0,1,..,2l,
\eea
where $A^{mn}$ is square matrix of the form
\bea\label{matrix}
A^{mn}=
\left(
\begin{array}{ccccccc}
0&-2l\gamma^2&0&0&...&0&0
\\[2pt]
1&0&-(2l-1)\gamma^2&0&...&0&0
\\[2pt]
0&2&0&-(2l-2)\gamma^2&...&0&0
\\[2pt]
0&0&3&0&...&0&0
\\
\vdots&\vdots&\vdots&\vdots&\ddots&\vdots&\vdots
\\[2pt]
0&0&0&0&\cdots&-2\gamma^2&0
\\[2pt]
0&0&0&0&\cdots&0&-\gamma^2
\\[2pt]
0&0&0&0&\cdots&2l&0
\end{array}
\right),
\eea
accommodates the $l$-conformal Galilei symmetry. The corresponding symmetry transformations involve (\ref{transform}) and \cite{GM}
\begin{multline}
{x'}_{n,i}(t)=x_{n,i}(t)-(a+bt^2+ct)\dot{x}_{n,i}(t)-\omega_{ij}x_{n,j}(t)+
\\[2pt]
+\rho^{2(n-l)}\sum_{s=0}^{n}\sum_{m=s}^{2l}\frac{(-1)^{n-s}m!(2l-s)!}{s!(m-s)!(n-s)!(2l-n)!}t^{m-s}
\left(\frac{\dot{\rho}}{\rho}\right)^{n-s}\lambda_{m,i}.
\end{multline}

The matrix (\ref{matrix}) has the following eigenvalues \cite{GM}
\bea\label{eig}
\begin{aligned}
&
0,\,\pm 2i\gamma,\,\pm 4i\gamma,\,...,\,\pm2il\gamma &&\mbox{- for integer $l$};
\\[2pt]
&
\pm i\gamma,\,\pm3i\gamma,\,\pm5i\gamma,...,\,\pm2il\gamma &&\mbox{- for half-integer $l$}.
\end{aligned}
\eea
Let us denote the eigenvector which corresponds to eigenvalue $ia\gamma$ as $\vartheta_{a,n}$, $a=-2l,-2l+1,..,2l$. Then, according to (\ref{equ1}), the dynamics of the field $\sum\limits_{n=0}^{2l}x_{n,i}\vartheta_{a,n}$ is governed by the equation
\bea
(\mathbb{D}-ia\gamma)\sum\limits_{n=0}^{2l}x_{n,i}\vartheta_{a,n}=0.
\nonumber
\eea
This fact allows one to obtain fields which are described by the fourth-order differential equation. Indeed, for each pair of eigenvalues $ia\gamma$, $ib\gamma$ ($a\neq \pm b\neq 0$), the dynamics of the real field $\sum\limits_{n=0}^{2l}x_{n,i}(\vartheta_{a,n}+\bar{\vartheta}_{a,n}+\vartheta_{b,n}+\bar{\vartheta}_{b,n})$ is satisfied to
\bea\label{dr1}
(\mathbb{D}^2+a^2\gamma^2)(\mathbb{D}^2+b^2\gamma^2)\sum\limits_{n=0}^{2l}x_{n,i}(\vartheta_{a,n}+\bar{\vartheta}_{a,n}+
\vartheta_{b,n}+\bar{\vartheta}_{b,n})=0,
\eea
where the bar stands for complex conjugate. The system of equations (\ref{1}a) and (\ref{dr1}) enjoys the $l$-conformal Galilei symmetry.

For integer $l$, the third-order differential equations can be generated by using the eigenvector corresponding to the zero eigenvalue. For example, the dynamics of the real field $\sum\limits_{n=0}^{2l}x_{n,i}(\vartheta_{0,n}+\vartheta_{a,n}+\bar{\vartheta}_{a,n})$ is described by
\bea
(\mathbb{D}^3+a^2\gamma^2\mathbb{D})\sum\limits_{n=0}^{2l}x_{n,i}(\vartheta_{0,n}+\vartheta_{a,n}+\bar{\vartheta}_{a,n})=0.
\nonumber
\eea

For the order higher than four, differential equations, which possess the $l$-conformal Galilei symmetry, can be obtained by analogy. Let $ia_1\gamma$, $ia_2\gamma$,.., $i a_s\gamma$ be the eigenvalues of the matrix (\ref{matrix}), where $a_1$, $a_2$,.., $a_s$ are distinct positive numbers. Then the evolution of the real field $\chi_i=\sum\limits_{n=0}^{2l}x_{n,i}\sum\limits_{k=1}^s(\vartheta_{a_k,n}+\bar{\vartheta}_{a_k,n})$ is satisfied to the following $2s$-order differential equation:
\bea\label{dr2}
\prod_{k=1}^{s}(\mathbb{D}^2+a_k^2)\chi_i=0\,\rightarrow\,\chi_i=\sum_{k=1}^{s}[A_{k,i}\cos{(a_k\, s(t))}+B_{k,i}\sin{(a_k\, s(t))}],
\eea
where $A_{k,i}$, $B_{k,i}$ are constants of integration.

For integer $l$, one constructs the odd-order differential equations along similar lines. In particular, one may obtain the following $(2s+1)$-order equation
\bea\label{dr3}
\prod_{k=1}^{s}(\mathbb{D}^2+a_k^2)\mathbb{D}\psi_i=0\,\rightarrow\,\psi_i=A_{0,i}+\sum_{k=1}^{s}[A_{k,i}\cos{(a_k\, s(t))}+B_{k,i}\sin{(a_k\, s(t))}],
\eea
where the field $\psi_i$ is defined by
\bea
\psi_i=\sum\limits_{n=0}^{2l}x_{n,i}\left(\vartheta_{0,n}+
\sum\limits_{k=1}^s(\vartheta_{a_k,n}+\bar{\vartheta}_{a_k,n})\right).
\nonumber
\eea
The remarkable property of the particles, which described by (\ref{dr2}), (\ref{dr3}), is that their motion is bounded.

According to (\ref{eig}), one and the same eigenvalues of the matrix (\ref{matrix}) appear for different values of parameter $l$. Therefore, one can realize different $l$-conformal Galilei algebras in one and the same system of equations. For example, the system (\ref{1}a), (\ref{PU4}) may accommodate $l$-conformal Galilei symmetry for any integer $l>2$. It should be stressed that a number of functionally independent constants of the motion, which correspond to the vector generators in the algebra, correlates with the order of obtained differential equation (see also a related discussion in \cite{GM}).

\vskip 0.5cm
\noindent
{\bf{\large 4. Conclusion}}
\vskip 0.5cm

To summarize, in this work the way to construct stable higher-derivative mechanical model, which enjoys $l$-conformal Galilei symmetry, has been introduced. The dynamics of the model is characterized by two differential equations. The first one can be viewed as describing an effective external field. The second equation can be interpreted as corresponding to the PU oscillator moving in this external field. The obtained system of equations is not Lagrangian. The way to derive conserved charges associated with the symmetry transformations of the model has been discussed. It has been observed that a number of functionally independent integrals of motion, which correspond to the vector generators in the algebra, correlates with the order of the PU oscillator-like equation.

Turning to further possible developments, it would be interesting to find such a set of auxiliary fields which provide a Lagrangian formulation for the model constructed in the present work. It is also worth to construct the Newton-Hooke counterpart of the model introduced in this paper. In this context the results obtained in Ref. \cite{Galajinsky_1} may be useful. These issues will be studied elsewhere.

\vskip 0.5cm
\noindent
{\bf{\large Acknowledgements}}
\vskip 0.5cm
We thank A. Galajinsky for the comments on the manuscript and S.O. Krivonos for his interest in the work. This work was supported by the MSE program Nauka under the project 3.825.2014/K, and RFBR grant 15-52-05022.

\fontsize{10}{13}\selectfont

\end{document}